\documentclass[english,aps,prb,floatfix,twocolumn,showpacs]{revtex4}
\usepackage{amsmath}
\usepackage{babel}
\usepackage{graphicx}
\usepackage{amssymb}
\usepackage{multirow}
\usepackage{color}

\begin{document}

\ifcsname DIFaddbegin\endcsname%
  \newcommand{\kh}[1] {\DIFaddend\textcolor{green}{{[}KH: #1{]}}\DIFaddbegin}
  \newcommand{\mw}[1] {\DIFaddend\textcolor{green}{{[}MW: #1{]}}\DIFaddbegin}
\else
  \newcommand{\kh}[1] {}
  \newcommand{\mw}[1] {}
\fi

\title{Magnetism in Sr$_{2}$CrMoO$_{6}$: A Combined Ab-initio and Model Study}
\author{Prabuddha Sanyal$^{1}$, Anita Halder$^{2}$, Liang Si$^{3}$, Markus Wallerberger$^{3}$, Karsten Held$^{3}$ and Tanusri Saha-Dasgupta$^{2}$} 
\affiliation{$^{1}$ Department of Physics, Indian Institute of Technology, Roorkee 247667, India}
\affiliation{$^{2}$ Department of Condensed Matter Physics and Materials Science, S.N. Bose National Center for Basic Sciences, Kolkata,India}
\affiliation{$^{3}$ Institute for Solid State Physics, TU Wien, 1040 Wien, Austria}

\date{\today}

\begin{abstract}
Using a combination of first-principles density functional theory (DFT) calculations and exact diagonalization studies of a 
first-principles derived model, we carry out a microscopic analysis of the magnetic properties of the half-metallic double perovskite 
compound, Sr$_2$CrMoO$_6$, a sister compound of the much discussed material Sr$_2$FeMoO$_6$.
The electronic structure of Sr$_2$CrMoO$_6$, though appears similar to Sr$_2$FeMoO$_6$ 
at first glance, shows non trivial differences with that of Sr$_2$FeMoO$_6$ on closer examination. 
\textcolor{black}{In this context, our study highlights the importance of charge transfer energy between the two 
transition metal sites. The change in charge transfer energy due to shift of Cr $d$ states in 
Sr$_2$CrMoO$_6$ compared to Fe $d$ in Sr$_2$FeMoO$_6$ suppresses the hybridization between 
Cr $t_{2g}$ and Mo $t_{2g}$.} This strongly weakens the 
hybridization-driven mechanism of magnetism discussed for 
Sr$_2$FeMoO$_6$. Our study reveals that, nonetheless,
the  magnetic transition temperature of  Sr$_2$CrMoO$_6$ remains high 
since additional superexchange contribution to magnetism
arises with a finite intrinsic moment developed at the
Mo site. \textcolor{black}{We further discuss the situation in comparison to another related double perovskite 
compound, Sr$_2$CrWO$_6$.} We also examine the effect of correlation beyond DFT, using dynamical mean field theory (DMFT). 

\end{abstract}

\maketitle
\noindent
\section{Introduction}

In recent years, double perovskites with general formula A$_2$BB$^{'}$O$_6$ (A: rare-earth/alkaline-earth cation, 
B: 3$d$ transition metal, B$^{'}$: 4$d$/5$d$ transition metal)  have been \textcolor{black}{in focus of discussion
due to their attractive properties. This includes multiferrocity,\cite{bimno,recent1} magnetodielectric and magnetocapacitive
properties,\cite{lmno,lcmo} complex spin behavior,\cite{recent-spin} magneto-structural coupling\cite{recent2} etc.
Many of these compounds show half-metallic behavior, which is of
importance for spintronics and related technological applications.~\cite{SaitohPRB,ES,tomioka,TSD,Mag-Res1,Mag-Res2,Mag-Res3,Saitoh9}
The  high magnetic transition temperatures ($T_{c}$) exhibited by compounds like Sr$_{2}$FeMoO$_{6}$, open up
the possibility of room temperature application.\cite{SFMO,DDreview} 
The pioneering work, reporting\cite{SFMO} room temperature magnetoresistivity of Sr$_2$FeMoO$_6$ (SFMO) was met with
excitement and followed by a surge of activity
exploring the dependence of 3d transition metal (TM) ion on the properties of these compounds.}

Following this motivation,  Cr-based double perovskites have been synthesized and studied,\cite{cr} in particular, the sister compound of SFMO, Sr$_{2}$CrMoO$_{6}$ (SCMO).\cite{inorg} 
Unlike in SFMO, there can be no valence compensation between Cr and the Mo  in SCMO: Cr can only be
in the 3+ state making Cr$^{3+}$/Mo$^{5+}$ only possible combination, while for SFMO, both Fe$^{3+}$/Mo$^{5+}$ 
and Fe$^{2+}$/Mo$^{6+}$ combinations are possible. Thus, SCMO was expected to be an even better candidate for room temperature spintronics.                
However, although the measured transition temperature of SCMO is
high,\cite{cnr,inorg} the observed moment\cite{prb} and the magnitude of the tunneling magnetoresistance 
turned to be disappointingly low. The presence of large antisite disorder, together with oxygen vacancy was held responsible 
for this.\cite{cnr} The antisite disorder present in the samples was estimated to be as high as 43 - 50 $\%$.\cite{inorg,disorder} To appreciate 
the role of 3$d$ TM in the properties of double
perovskites, it is thus highly desirable to understand the electronic and magnetic properties of pure SCMO.

Theoretical studies on SCMO have been carried out within the framework of DFT.\cite{physica,Wu} However no microscopic analysis has been carried out, neither has the transition temperature been
 calculated, except for  a recent mean field analysis \cite{ngantso}
based on a classical Ising model. The latter neglects the itinerant electronic character of Mo 4$d$ electrons, a crucial
component in understanding the behavior of these 3$d$-4$d$ double perovskites.

\textcolor{black}{Futhermore, though the hybridization driven mechanism of magnetism has been identified\cite{TSD} a while ago as the driving force 
in setting up the high Curie temperature in compounds like SFMO, the role of charge transfer between B and B$^{'}$ sites in the 
perspective of properties of double perovskite compounds in general, has not been studied in detail.
In particular no systematic study exists. This is however an important issue in understanding the comparative magnetic properties of double perovskite compounds.}

In the present study, we aim to fill this gap  by combining state-of-art DFT calculations and exact diagonalization  of the DFT derived 
model Hamiltonian. \textcolor{black}{We consider the case of SCMO, in comparison to SFMO as well as Sr$_{2}$CrWO$_{6}$ (SCWO)\cite{cr} which is another 
double perovskite from the Cr family but with 5$d$ W instead of 4$d$ Mo.} Our microscopic study shows a considerable suppression of the Cr $t_{2g}$ - Mo $t_{2g}$ hybridization in SCMO compared 
to that in SFMO \textcolor{black}{or SCWO}, driven by the change in charge transfer energy between B and B$^{'}$ sites in different compounds. The suppressed hybridization in SCMO makes the 
Mo $t_{2g}$ electrons more localized compared to SFMO \textcolor{black}{or SCWO}. This, in turn, opens  up an additional, super-exchange contribution to magnetism. The calculated
magnetic transition temperature $T_c$ without super-exchange contribution, gives about 79 $\%$ of the measured $T_c$. 
The consideration of additional super-exchange contribution makes the calculated $T_c$ comparable to the measured one. Our work reveals the importance of the super-exchange in SCMO in addition to the hybridization-driven mechanism operative in SFMO \textcolor{black}{or SCWO, and underlines the crucial role played by the charge transfer energy in activating the
additional superexchange channel in SCMO.\cite{SFMO}}

{We further study the effect of correlation on the half-metallic property of SCMO by including local correlations on top of DFT
within the framework of DMFT. The DMFT results confirm the
half-metallic ground state below the magnetic transition temperature, implying that the qualitative description  remains 
unaffected by
 correlation effects.}

\section{Computational Details}
The first-principles DFT calculations have been carried out using the plane-wave pseudopotential method implemented
within the Vienna Ab-initio Simulation Package (VASP).\cite{VASP} The exchange-correlation functional was considered within the generalized 
gradient approximation (GGA) \textcolor{black}{in the framework of PW91.\cite{gga} Additionally calculations have been carried
with exchange-correlation functional of local density approximation (LDA)\cite{lda} as well as GGA within the framework of 
PBE,\cite{pbe} in order to check any possible influence of the  exchange-correlation functional on the calculated electronic structure.} 
The projector-augmented wave (PAW) potentials\cite{paw} were used and the wave functions 
were expanded in the plane-wave basis with a kinetic-energy cutoff of 500 eV. Reciprocal-space integrations were carried out with 
a $k$-space mesh of 8 $\times$ 8 $\times$ 8.

In order to extract a few-band tight-binding (TB) Hamiltonian out of the full DFT calculation which can be used as input to multiorbital, 
low-energy model Hamiltonian in exact diagonalization calculations we have carried out $N$-th order muffin tin orbital
($N$MTO) calculations.\cite{nmto} A prominent feature of this method is the downfolding scheme. Starting from a full DFT
calculation, it defines a few-orbital Hamiltonian in an energy-selected, effective Wannier function basis, by integrating out the 
degrees of freedom that are not of interest. The NMTO technique relies on the self-consistent potential parameters
obtained out of linear muffin-tin orbital (LMTO)\cite{lmto} calculations. In order to cross-check the TB parameters generated 
out of NMTO-downfolding calculations, further calculations were carried out using wien2wannier \cite{wien2wannier}.
This generates maximally
localized Wannier functions\cite{Wan90} from Wien2K\cite{lapw} which employs a full potential linear augmented plane wave (FLAPW) basis. For self-consistent DFT calculation in the
FLAPW basis the number of $k$-points in the irreducible Brillouin zone was chosen to be 64.
The commonly used criterion relating the plane wave and angular momentum cutoff, 
$l_{max}$ = $R_{MT} \times K_{max}$ was chosen to be 7.0, where R$_{MT}$ is the smallest MT sphere
radius and $K_{max}$ is the plane wave cutoff for the basis. The chosen atomic radii for Sr, Cr, Mo and O were 
1.43 \AA, 1.01 \AA, 1.01 \AA, and 0.87 \AA, respectively.
 
The exact diagonalization of the ab-initio derived low-energy Hamiltonian, defined in the Wannier function basis 
has been carried out for finite-size lattices of dimensions 4$\times$4$\times$4, 6$\times$6$\times$6, and 8$\times$8$\times$8.
The results presented in the following are for 8$\times$8$\times$8 lattice.

The DFT+DMFT calculations\cite{dmft1,dmft2} have been carried out using the wien2wannier\cite{wien2wannier}-derived maximally localized Wannier functions of Wien2K 
as a starting point.
We restricted the DMFT to the low energy degrees of freedom, i.e., to the
``$t_{2g}$'' orbitals of Mo and the ``$t_{2g}$'' and ``$e_g$'' orbitals of Cr
(These low energy
orbitals are actually a mixture between predominately transition metal $t_{2g}$ ($e_g$) character with some admixture of oxygen $p$ character).
We supplemented the DFT low-energy Hamiltonian in the Wannier basis by a local Coulomb interaction in  Kanamori parametrization. For details 
see Ref.\ \onlinecite{w2d}. For the interaction values we chose the typical values for $3d$ and $4d$ transition metal oxides. 
The choices were: interorbital Coulomb repulsion of
$U'=4\,$eV (2.4 eV) and a Hund's coupling $J_H=0.7\,$eV (0.3 eV) for Cr (Mo) as estimated for neighboring vanadium \cite{Taranto2013}(ruthenium\cite{Si2015}) 
perovskites.  The intraorbital (Hubbard) repulsion follows from orbital symmetry as $U=U'+2J_H$; and the pair hopping term is of equal strength as $J_H$. As a DMFT impurity solver, continuous-time quantum Monte-Carlo simulations\cite{CTQMC} in the w2dynamics\cite{w2d} implementation was used 
which includes the full SU(2) symmetry.  The  effect of electronic correlations beyond GGA, within the framework of
static theory was also checked by performing GGA+$U$ calculations with choice of same $U$ parameters, as in DMFT calculation.
\textcolor{black}{We have also checked the validity of our results by varying the U value by +/-1 eV at Cr site, and by +/-0.5 eV at Mo site. 
The trend in the results was found to remain unchanged.}

\section{Results}

\subsection{Basic DFT Electronic Structure}

We first revisit the basic DFT electronic structure of SCMO, which has been calculated before using variety of basis sets, including 
plane wave,\cite{jpc} LAPW\cite{physica} and LMTO.\cite{Wu} SCMO crystallizes in the cubic Fm3m space group (225), with lattice parameter of
7.84 \AA.\cite{cnr} 

The upper panel of Fig 1. shows the spin polarized density of states (DOS) calculated in a plane wave basis. The basic features of the DOS, which agree
very well with previous studies\cite{jpc,physica,Wu} are the following: The states close
to Fermi level, E$_F$, (set as zero in the figure), are dominated by Cr and Mo $d$ states hybridized with O $p$ states. 
The states of dominant O $p$ character are positioned further down in energy, separated by a small gap from Cr and Mo $d$ states.
Empty Sr states, not shown in the energy scale of the figure, remain far above E$_F$. Due to the large octahedral crystal field
at Mo sites, the empty Mo $e_g$ states lie far above E$_F$. The Cr $t_{2g}$ states are occupied in the \textcolor{black}{up} spin channel
and empty in the \textcolor{black}{down} spin channel. Cr $e_g$ states are empty in both spin channels, in agreement with the nominal Cr$^{3+}$
($d^{3}$) valence. The empty and highly peaked Mo $t_{2g}$ states in the \textcolor{black}{up} spin channel appear in between the crystal field split
Cr $t_{2g}$ and Cr $e_g$ states for the same spin. In contrast for the \textcolor{black}{down} spin channel, the  Mo $t_{2g}$ hybridize more strongly  with Cr $t_{2g}$, which explains their much larger bandwidth.

The above described basic features for B and B$^{'}$ states, are rather similar for SCMO and SFMO,\cite{la-sfmo} 
and in that respect also for SCWO.\cite{3d-5d} Nevertheless, the DOS of SCMO differs from both SFMO and SCWO in important details. The hybridization between B $t_{2g}$ and B$^{'}$ $t_{2g}$ in the \textcolor{black}{down} spin channel, is found to be significantly lower compared
to that of SFMO or SCWO.\cite{la-sfmo,3d-5d} A measure of this is 
the Cr contribution in the \textcolor{black}{down} spin bands crossing E$_F$ of predominant Mo character. We estimate this contribution or admixture to be 35$\%$ for SCMO, 
while the corresponding estimates for SCWO and SFMO are much larger, 66$\%$ and 72 $\%$, respectively.

The top rows in Table I show the calculated magnetic moments at Cr and Mo sites, as well as the total moment \textcolor{black}{in three choices
of exchange-correlation functionals, GGA-PW91, GGA-PBE as well as LDA.  The half-metallic nature of the ground state is found to be robust
in all different choices of exchange-correlation functionals with the total magnetic moment of 2.0 $\mu_B$/f.u. and finite moments residing at O sites. As is seen, while the individual moments on Cr and Mo are found to vary depending on the nature of approximation, the magnetic moments on 
Cr and Mo site being aligned in an antiparallel manner, the enhancement/reduction of individual moments cancel, thereby retaining the 
half-metallicity and net moment of  2 $\mu_B$/f.u.} Our calculated GGA
magnetic moments at Cr and Mo sites are in good agreement with that reported by Liet {\it et al.}\cite{jpc}
The calculated site-specific moment reported by Wu \cite{Wu} is however much smaller than that of our as well as that 
of Liet {\it et al.}\cite{jpc} 
This is presumably due to different choices of the muffin tin radii as well as the exchange-correlation functional. 

The measured magnetic moments are significantly smaller than the  theoretical values. Experimentally, the total moment is only 0.5 $\mu_B$\cite{prb} 
and the  moment on the Cr sites is 0.8 $\mu_B$.\cite{sss} This discrepancy was hitherto argued to be due to the large disorder present in the sample.\cite{cnr} 
It is interesting to note that the value of the calculated GGA-PW91 magnetic moment at the Mo sites (0.43 $\mu_B$) is larger than the calculated B$^{'}$ site moment 
for SFMO (0.23 $\mu_B$) and  SCWO (0.30 $\mu_B$). This indicates a weaker itinerancy of the Mo electrons  in SCMO, compared to  SFMO or to that of W in SCWO. 
This in turn suggests that a small, but finite intrinsic moment develops at the Mo site as a consequence of the weaker hybridization 
between Cr and Mo. The suppression of the hybridization and the reduced itinerancy of the Mo $t_{2g}$ electrons in SCMO has
been also pointed out in the study by Wu,\cite{Wu} though no detailed understanding of the mechanism of magnetism
was provided. In the following we will elaborate on the microscopic mechanism of magnetism in SCMO.

\begin{figure}[h]
\centerline{\includegraphics[width=7cm,keepaspectratio]{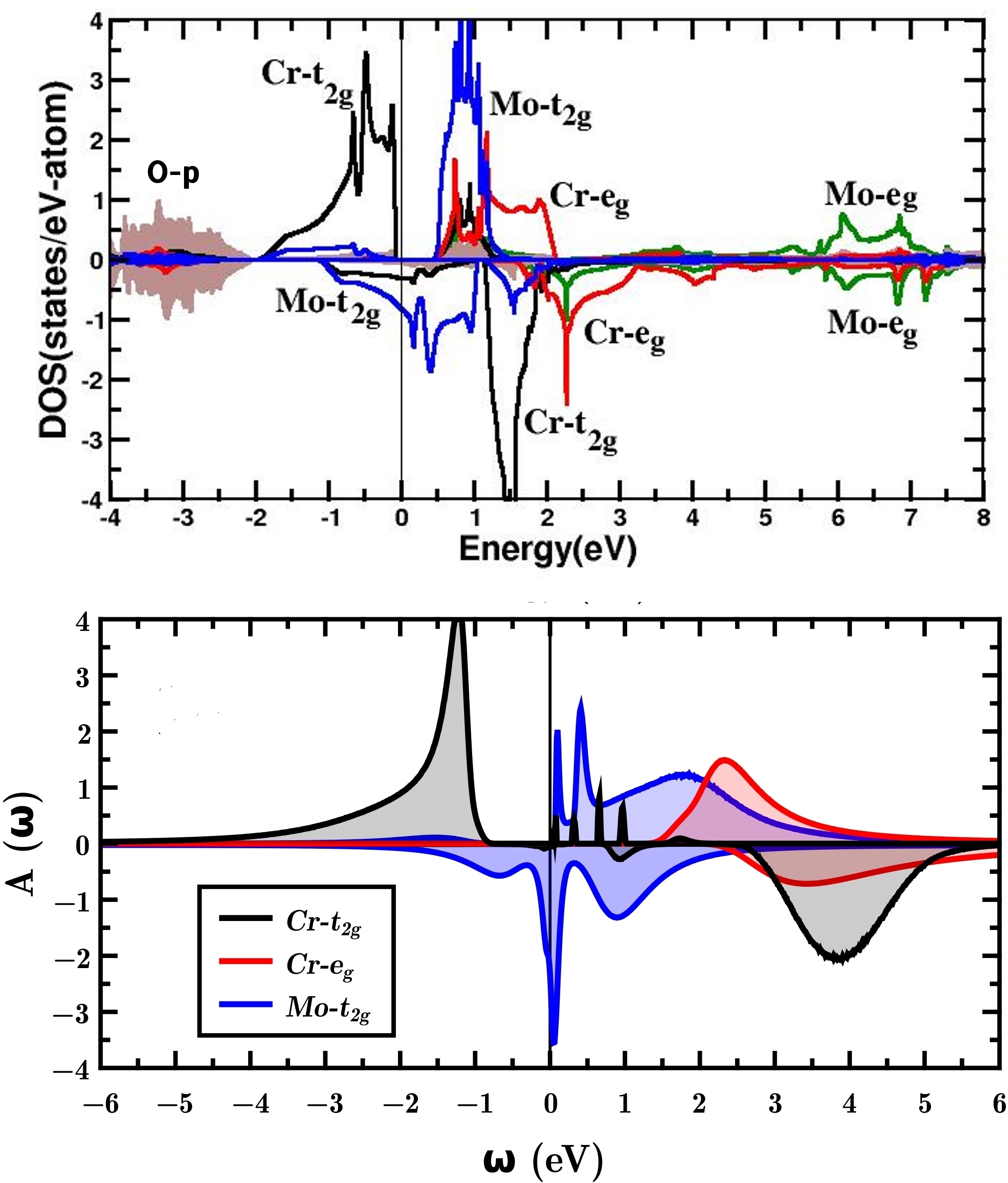}}
\caption{(Color online) Upper panel: GGA-PW91 spin-polarized DOS projected onto Cr $t_{2g}$ (black solid line), Cr $e_g$ (red solid line), 
Mo $t_{2g}$ (blue solid line), Mo $e_g$ (green solid line) and
O p (shaded area) in the FLAPW basis. The zero of the energy is set to the Fermi energy.
\textcolor{black}{The positive (negative) axis of DOS corresponds to DOS in up (down) spin channel.}  
Lower panel: DMFT spectral density, calculated at 200 K within the 
Mo $t_{2g}$ and Cr $t_{2g}+e_{g}$ Wannier basis.\cite{41}}
\end{figure}

\textcolor{black}{
\begin{table}
\begin{tabular}{|ccccc|}
\hline
 & Sr & Cr & Mo & total \\
\hline
LDA& 0.00 & -2.18 & 0.32 & -2.00 \\
GGA-PW91 & 0.00 & -2.29 & 0.43 & -2.00 \\
GGA-PBE & 0.00 & -2.36 & 0.49 & -2.00 \\
GGA+$U$ & 0.00 & -2.54 & 0.58 & -2.00 \\
DMFT & 0.00 & -2.84 & 0.84 & -2.00 \\
\hline
\end{tabular}
\caption{Calculated magnetic moments (in $\mu_B$) within \textcolor{black}{LDA, GGA-PW91, GGA-PBE,} GGA+$U$ and DMFT. Note that within GGA there is also a moment on the oxygen sites which 
is accounted for in the Cr and Mo moment in DMFT as the predominately metal $d$ Wannier functions also have some oxygen admixture.}
\label{Table1}
\end{table}
}

\subsection{DMFT Spectral Density} 

In order to study the effect of electronic correlations, specially the dynamical correlation which may be important for SCMO due to the metallic nature of the ground state, we further carry out DMFT calculations in the Wannier function basis.
The lower panel in Fig 1, shows the DMFT spectral density calculated at a temperature of ~200 K.

The spectrum shows qualitatively similar behavior to that obtained in DFT calculation, though the \textcolor{black}{down} spin 
spectrum shows the correlation physics with a feature resembling lower and 
upper Hubbard bands and a well defined quasi-particle peak at Fermi energy. 
This demonstrates the dual nature of the Mo \textcolor{black}{down} spin
with localized electrons (Hubbard bands) and 
itinerant electrons (quasiparticle band) at  the same time.
The fully spin-polarized conducting electrons are  Fermi-liquid-like with a linear frequency-dependence of the self energy (not shown). Hence the quasiparticle peak will lead to a 
Drude peak in the optical conductivity. 

The DMFT magnetic moments are shown in the last row of Table I. We note the DMFT calculated magnetic moments are in the basis of Cr and Mo $t_{2g}$ Wannier functions, which includes the effect of oxygen. This makes a direct comparison between DMFT and GGA values 
somewhat more difficult. However since the difference between GGA and DMFT magnetic moments is larger than the GGA oxygen contribution, 
we can conclude that electronic correlations somewhat enhance the magnetic moment on both, Mo and Cr, sites. This is further
supported by the magnetic moments calculated within the static theory of GGA+$U$\textcolor{black}{, with choice of same $U$ parameters
as in DMFT calculation}, shown in the second row of Table I, which shows an enhancement of the moment both at Cr and Mo sites, compared to 
that of GGA.

\subsection{Few-orbital, low-energy Hamiltonian}

In order to understand the driving mechanism of magnetism in SCMO in a more quantitative manner, we carry out a $N$MTO downfolding 
in order to estimate the positions of exchange split Mo $t_{2g}$ energy levels, before and after switching on the hybridization between the Mo $t_{2g}$ and
Cr $t_{2g}$. The former provides the estimate of intrinsic spin splitting at the Mo site,
 while the latter
 provides the information of
the spin splitting at the Mo site renormalized by the hybridization effect from Cr $t_{2g}$. As a first step of this procedure, we downfold 
O $p$, Sr as well as Cr and Mo $e_g$ degrees of freedom. This defines a few-orbital Hamiltonian consisting of Cr $t_{2g}$ and Mo $t_{2g}$ 
states. In the second step, we perform a further downfolding, keeping only the Mo $t_{2g}$ degrees of freedom, i.e.,  downfolding everything else 
including the Cr $t_{2g}$ degrees of freedom. On site matrix elements of the few orbital Hamiltonian in real space representation defined in the 
Cr $t_{2g}$ - Mo $t_{2g}$ basis and the massively downfolded basis give the energy level positions before and
after switching of the hybridization, respectively. The obtained result is presented in Fig 2. First of all, we notice that the Mo $t_{2g}$ states
are energetically situated in between the exchange-split energy levels of Cr $t_{2g}$'s. Thus switching of the hybridization between
Cr and Mo, pushes the Mo \textcolor{black}{up} spin states down because these  are below the Cr states of the same spin. In contrast, the Mo \textcolor{black}{down} spins 
are above their Cr counterpart and hence the hybridization shifts them upwards.
Thus the hybridization induces a renormalization of the  spin splitting at the Mo site, with renormalized value of about 0.70 eV, and being
oppositely oriented (negative) with respect to that at Cr site. \textcolor{black}{Note the intrinsic (in absence of hybridization) spin splitting at Mo site is small, having a value of 0.15 eV.}

In this respect, the situation is very similar to SFMO or SCWO, for which also a negative
splitting is induced at the itinerant B$^{'}$ sites because of the hybridization with the large spin at the 
B sites.\cite{TSD,SFMO,3d-5d,new} This supports that the hybridization-driven mechanism is operative in SCMO as well. 
While in all three cases of SFMO, SCWO and SCMO, the B$^{'}$ $t_{2g}$ states appear in between the strong
exchange split energy levels of B site, which is an essential ingredient for hybridization-driven mechanism
to be operative, we notice the relative energy position of B$^{'}$  $t_{2g}$ states with respect to the exchange
split B states is different in case of SCMO, as compared to SFMO or SCWO. For SFMO or SCWO, the \textcolor{black}{down} spin
B$^{'}$ $t_{2g}$ states appear very close to B site \textcolor{black}{down} spin states, while for SCMO, they are shifted down.\cite{3d-5d,SFMO}
This hints towards a significantly different charge transfer energy in case of SCMO as compared to SFMO or SCWO.
This will be elaborated in the next paragraphs.
We further notice that the intrinsic splitting at the Mo sites (0.15 eV) is somewhat larger than for
 SCWO (0.05 eV).\cite{3d-5d}
This further confirms the conclusion drawn from the calculated magnetic moment at Mo site that is in SCMO, unlike SFMO, Mo has 
a finite intrinsic moment. The magnetism in SCMO, as mentioned already, 
thus has an additional contribution, originating
from the superexchange between the large moment at the Cr site and the intrinsic moment 
at the Mo site, on top of the
hybridization-driven mechanism, as operative in SFMO. Note that the superexchange is antiferromagnetic, 
aligning the Cr and Mo moments antiparallely, {\it i.e.}, in the same way as for 
the hybridization-driven mechanism.

\begin{figure}[h]
\centerline{\includegraphics[width=7cm,keepaspectratio]{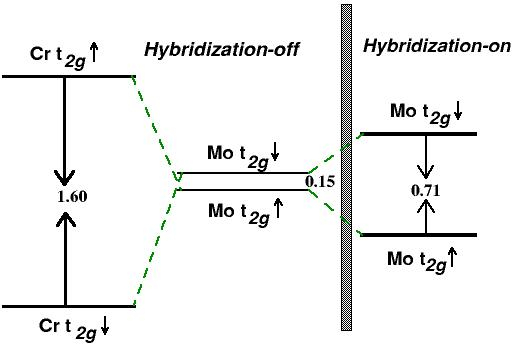}}
\caption{(Color online) The energy level diagram for SCMO, in the absence and presence of the Cr-Mo hybridization. 
The energies are in units of eV.}
\end{figure}

In this situation, the low-energy model Hamiltonian for SCMO in the 
 Cr $t_{2g}$ and Mo $t_{2g}$ Wannier basis, describing the hybridization and super-exchange mechanism is given by \cite{3d-5d,la-na}
\begin{eqnarray}
H & = & \epsilon_{Cr}\sum_{i\in B}f_{i\sigma\alpha}^{\dagger}f_{i\sigma\alpha}+
\epsilon_{Mo}\sum_{i\in B'}m_{i\sigma\alpha}^{\dagger}m_{i\sigma\alpha}  \nonumber
\\ \nonumber
& & -t_{Cr-Mo}\sum_{<ij>\sigma,\alpha}f_{i\sigma,\alpha}^{\dagger}m_{j\sigma,\alpha} 
 \\ \nonumber
& & -t_{Mo-Mo}\sum_{<ij>\sigma,\alpha}m_{i\sigma,\alpha}^{\dagger}m_{j\sigma,\alpha} \\ \nonumber
& & -t_{Cr-Cr}\sum_{<ij>\sigma,\alpha}f_{i\sigma,\alpha}^{\dagger}f_{j\sigma,\alpha} 
+ J\sum_{i\in Cr} {\bf S}_{i} \cdot
f_{i\alpha}^{\dagger}\vec{\sigma}_{\alpha\beta}f_{i\beta} \\
& & + J^{'}\sum_{i\in Cr,j\in Mo} {\bf S}_{i} \cdot {\bf s}_{j} \; .
\label{fullham}
\end{eqnarray}

Here the $f$'s and $m$'s are second quantization operators for the Cr $t_{2g}$ and Mo $t_{2g}$ degrees of freedoms; 
$\sigma$ is the spin index and $\alpha$ is the orbital index that spans the $t_{2g}$ 
manifold;
$t_{Cr-Mo}$, $t_{Mo-Mo}$, $t_{Cr-Cr}$ represent the nearest neighbor Cr-Mo, the second
nearest neighbor Mo-Mo and the Cr-Cr hopping, respectively.   The on-site energy difference between
Cr $t_{2g}$ and Mo $t_{2g}$ levels is 
${\Delta} = \epsilon_{Cr} - \epsilon_{Mo}$.
To take into account the dual nature of the electrons that are both, itinerant and localized,  
we include a  large core spin ${\bf S}_i$ at the Cr site that couples with the itinerant electron 
delocalized over the Cr-Mo network, via a double-exchange like mechanism.
The last term represents the superexchange mechanism in terms of 
coupling between the Cr spin (${\bf S}_i$) and the intrinsic moment on the Mo site ($s_{j}$). 

All TB parameters of the model Hamiltonian, i.e.,
$\Delta$, $t_{Cr-Mo}$, $t_{Mo-Mo}$, $t_{Cr-Cr}$, are extracted from a non-spin-polarized DFT calculations through 
two independent means: a) through the $N$MTO downfolding technique, and b) through the  construction of maximally localized
Wannier functions in the  basis of the effective Cr and Mo $t_{2g}$ degrees of freedom. {The latter scheme has been also employed 
in the DMFT study, presented before.} In both cases we 
integrate out all other degrees of freedom except for the Cr and Mo $t_{2g}$'s states. The comparison of the
full DFT non spin-polarized band structure and the few orbital TB bands in maximally localized Wannier function basis
is shown in Fig 3. The agreement between the two is found to be as good as possible, proving the effectiveness of the 
Wannier function projection.

\begin{figure}[h]
\centerline{\includegraphics[width=7cm,keepaspectratio]{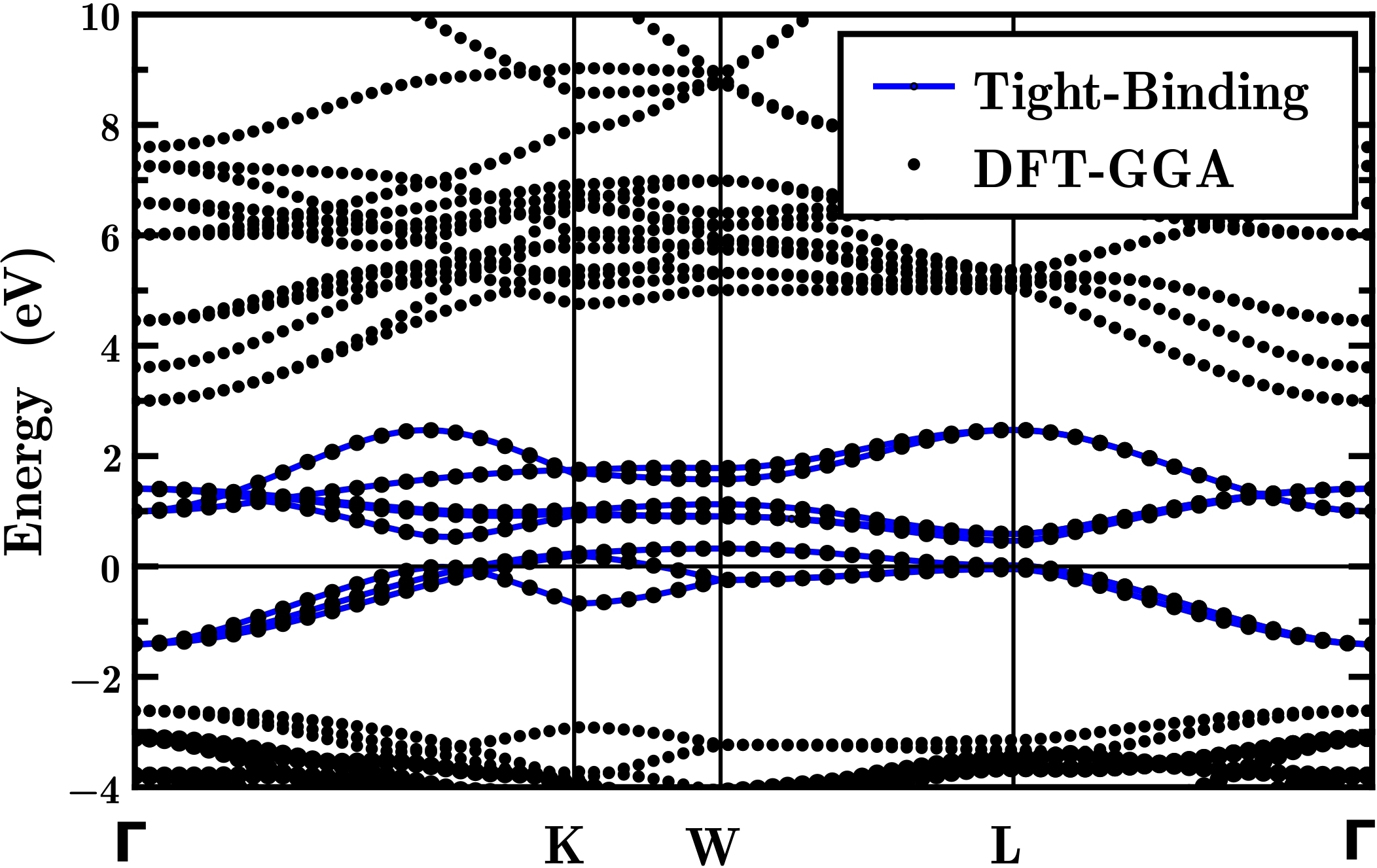}}
\caption{(Color online) Comparison of the full paramagnetic DFT band structure (solid line) and the few orbital TB band structure (+)
plotted along a high symmetry path through the  Brillouin zone [$\Gamma$: (0 0 0), K:   (3/8,3/4,3/8),
W:  (1/4,3/4,1/2), L:   (1/2,1/2,1/2)  and X:   (0,1/2,1/2)].}
\end{figure} 

The DFT estimates of $\Delta$, $t_{Cr-Mo}$,  $t_{Cr-Cr}$ and $t_{Mo-Mo}$, obtained by two different scheme of calculations
are shown in Table II. The agreement of the values in two independent scheme of calculations is remarkable and 
provides confidence regarding the results. \textcolor{black}{The TB parameters for SCWO\cite{3d-5d} and SFMO\cite{la-sfmo} 
are also listed in the table for comparison.}

\begin{table}
\begin{tabular}{|ccccc|}
\hline
   & $\Delta$ & $t_{Cr-Mo}$ & $t_{Mo-Mo}$ & $t_{Cr-Cr}$ \\
\hline
SCMO ($N$MTO) & -0.35 & 0.33 & 0.14 & 0.08 \\
\hline
SCMO (Wannier90) & -0.42 & 0.30 & 0.12 & 0.06 \\
\hline
SCWO & -0.66 & 0.35 & 0.12 & 0.08 \\
\hline
SFMO & -1.04 & 0.26 & 0.11 & 0.04 \\
\hline
\end{tabular}
\caption{Tight-binding parameters (in eV) of the few orbital Hamiltonian for SCMO, \textcolor{black}{SCWO\cite{3d-5d}, and SFMO\cite{la-sfmo}} 
in Wannier function basis, extracted out of DFT calculations.}
\label{TBparam}
\end{table}

While the estimated hopping parameters of SCMO are rather similar to that reported for SCWO,\cite{3d-5d} we find the charge transfer energy $\Delta$ (-0.4 eV) to be 
quite different from that estimated for SFMO (-1.0 eV)\cite{la-sfmo} and for SCWO (-0.7 eV).\cite{3d-5d}
\textcolor{black}{This forms the most crucial observation of our study, which dictates the differences in 
electronic structure of SCMO and, SFMO and SCWO. Related discussions involving importance of $\Delta$ and
$t$ in hybridization driven mechanism were mentioned by another group for comparisons between SFMO and
A-site B-site randomly doped Ti oxides although it was not a systematic one.\cite{Ti}}

The change in the bare,  non-spin-polarized charge transfer energy ($\Delta$), in turn, 
modifies the renormalized charge transfer energy  $\Delta_R$
between Cr and Mo in the \textcolor{black}{down} spin channel in the spin-polarized phase. 
{It is much larger in SCMO  (0.81 eV) compared to that in SCWO (0.51 eV),\cite{3d-5d} due to the  
moving down the column of the periodic table from the 4d element Mo in SCMO to the 5d element W 
in SCWO.}
As a consequence of the larger 
charge transfer splitting $\Delta_R$ the effect of the  
hybridization which is governed by $t^{2}_{C-Mo}/\Delta_R$ is significantly suppressed in SCMO, by about $63\%$ compared to the corresponding value 
for SCWO. 

{The parameter involving $J$ in (1) corresponds to the spin-splitting at the Cr site, while  $J^{'}$ is 
obtained from the enhanced intrinsic spin-splitting observed at Mo site in case of SCMO as compared to that for SCWO compound at the W site. \textcolor{black}{As discussed below, frustration effects make the convergence of total energy calculations 
in different spin configurations difficult, prohibiting direct extraction of $J^{'}$ in a single system. 
Thus $J^{'}$ needs to be determined by comparison of the independently obtained intrinsic spin splittings (obtained by turning off the B-B$^{'}$ hybridization) at the B$^{'}$ sites of the related compounds. This approach was used earlier,\cite{3d-5d}
where three compounds Sr$_2$CrWO$_6$, Sr$_2$CrReO$_6$ and Sr$_2$CrOsO$_6$ were considered. It was pointed out\cite{3d-5d} 
that the intrinsic splitting at the B$^{’}$ site due to a purely hybridization driven mechanism should increase proportional 
to the electron filling at B$^{'}$ site (the filling increases from 1 to 2 to 3 moving from W to Re to Os); the additional 
increase, if any, is due to the presence of the localized moment at B$^{'}$ site giving rise to superexchange 
$J^{'}$ between B and B$^{’}$ sites. In the present case, all the three compounds considered, namely SFMO, SCMO and SCWO have the same 
electron filling, namely 1. Hence considering the compound SCWO as the benchmark, the difference in intrinsic splitting 
between the Mo site of SCMO and the W site of SCWO is used to obtain the $J^{'}$ value for SCMO.\cite{footnote}} 

\subsection{Exact Diagonalization of Model Hamiltonian}

As the magnetism of SCMO turns out to be driven by both hybridization-driven mechanism and superexchange mechanism,
the calculation of magnetic transition temperature in a first-principles way is challenging. For example, an antiferromagnetic configuration of Cr spins leads to a frustration of the 
intrinsic Mo moment. On the other hand, the
hybridization-driven mechanism, disfavors the stabilization of magnetic configurations with Cr spins aligned in parallel to the 
spins at Mo sites. This thus leads to a frustration effect which makes the convergence of different spin configurations to extract
the magnetic exchanges a challenge within the framework of full blown DFT scheme. 

Hence, we consider the model Hamiltonian approach 
in the following. To this end, we solve the ab-initio constructed model Hamiltonian (1) using exact diagonalization on a finite 
lattice.  In particular, we consider the stability of the ferromagnetic arrangements of Cr spins 
measured as
the energy difference between the paramagnetic (PM) spin configuration and the ferromagnetic 
(FM) spin configuration. 
The PM  phase was simulated as disordered local moment calculations, where the calculations were carried out
for several ($\approx$ 50) disordered configurations of Cr spin and were averaged to get the energy corresponding to 
PM phase. This energy difference provides an estimate of the magnetic $T_c$.

In order to check the influence of the intrinsic moment at the Mo site on the magnetic transition, we first carried out exact diagonalization calculation 
considering the Mo site to be totally nonmagnetic, i.e., setting the last term of Hamiltonian (1) to zero. This  boils down to a hybridization-only driven mechanism of magnetism, as suitable for SFMO or SCWO.\cite{sfmo-model,3d-5d} 
The energy difference between the PM and FM in this calculation, turned out to be 0.067 eV/f.u., which is of the
same order as, but less than, that obtained using the TB parameters of SCWO (0.085 eV/f.u).\cite{3d-5d,note} As mentioned already,
the hopping parameters between SCMO and SCWO are very similar, thus the difference is caused by the different
charge transfer energy in the two compounds. Mapping this energy difference to the mean field transition temperature,
one would get $T_c$ in SCMO to be 79 $\%$ smaller compared to that of SCWO.\cite{cr}
The calculated PM and FM energy difference, considering the full model Hamiltonian, describing both hybridization-driven and 
super-exchange driven mechanism, is 0.080 eV/f.u., which makes the calculated $T_c$ of SCMO similar to that of SCWO.
{Mapping the PM and FM energy difference to the mean field $T_c$, one obtains the values 822 K for SCMO and 870 K for SCWO, which
are an overestimation of the experimental values.\cite{cr,inorg,cnr} This is presumably due to non-local fluctuations beyond mean-field.
 However, the trend is very well reproduced with T$_c^{SCMO}$/T$_c^{SCWO}$ = 0.95.}
  
\section{Summary}

Starting from a DFT description we provide a microscopic analysis of the magnetic behavior of SCMO, a sister 
compound of SFMO. DFT calculations on pure, defect-free SCMO show that, like SFMO and SCMO, it is a   half-metallic magnet. {The calculated DMFT results confirm the robustness of the half-metallicity of SCMO upon inclusion of dynamical correlation
effect.} The DMFT results show a splitting of the Mo \textcolor{black}{down} spin band into Hubbard bands and quasiparticle peak. This indicates the 
dual nature of the Mo electrons having both, a local spin moment and itinerant behavior.

The origin 
of magnetism in SCMO turns out to be somewhat different than in SFMO. The charge transfer energy between Cr 
and Mo in the \textcolor{black}{down} spin channel of SCMO is found to be larger than that between Fe and Mo in SFMO. This suppresses the effect of hybridization  between B and B$^{'}$ sites 
in SCMO compared to SFMO and has two consequences: (i) the hybridization-driven mechanism for 
magnetism is reduced and (ii)
  a small but finite intrinsic moment develops at the Mo sites. The latter gives rise to 
a partial localized character of the Mo electrons, which 
was absent in SFMO. This in turn opens up a super-exchange contribution to  magnetism in SCMO, which was absent for SFMO. We compare 
our results on SCMO to another Cr based double perovskite, SCWO. The magnetism in the latter, like SFMO,
is well described by a hybridization-only picture.
The computed $T_c$ obtained through exact 
diagonalization  of the ab-initio derived model Hamiltonian
show  the $T_c$ of SCMO to be similarly high as in SCWO,  only upon inclusion
of both, super-exchange and hybridization-driven contribution in case of SCMO. 
Thus while magnetism in both, SCWO and SFMO, is governed by hybridization, the story in SCMO 
appears with a twist. It needs to be described by a 
combination of hybridization and superexchange mechanism. 

We conclude that, in general, the magnetism in the double perovskite family has to be understood as an interplay between the hybridization
and super-exchange between B and B$^{'}$ sites. The relative contribution of one mechanism vs.\ the other one is dictated by the charge transfer
between the 3$d$ transition metal at the B site and the 4$d$ or 5$d$ transition metal at the B$^{'}$ site.

\section*{Acknowledgement}
PS acknowledges the SRIC project No PHY/FIG/100625. AH and TSD acknowledge support from DST through thematic Unit
of Execellence on Computational Materials Science; LS and KH from the European Research Council under the European Union's Seventh Framework Program (FP/2007-2013)/ERC through grant agreement n.\ 306447 and the  Austrian Science Fund (FWF) through the Doctoral School W1243 Solids4Fun (Building Solids for Function).
We are grateful to Tulika Moitra, P. Gunacker and D.D. Sarma for valuable discussions.


\begin{thebibliography}{99}
\bibitem{note} The energy difference between PM and FM phases for different lattice sizes 
are found to scale with size, although the ratio between
values for SCMO and SCWO is found to remain almost the same in calculations on different lattice sizes.
These values are found to be 0.050 eV/f.u. for SCMO,  0.063 eV/f.u. for SCWO with a ratio of 79.2 $\%$ for a 4x4x4 lattice,
0.066 eV/f.u. for SCMO,  0.084 eV/f.u. for SCWO with a ratio of 78.5 $\%$ for a 6x6x6 lattice, and
0.067 eV/f.u. for SCMO,   0.085  eV/f.u. for SCWO with a ratio of 78.8 $\%$ for a 8x8x8 lattice.
The values presented in the text are for 8x8x8.
\end{thebibliography}
\end{document}